\documentclass[12pt,letter]{article}

\usepackage{graphicx, epsfig, color}
\def\to{\rightarrow}

\def\bi{\begin{itemize}}
\def\ei{\end{itemize}}

\def\tb{\tilde b}
\def\tf{\tilde f}

\def\tst{\tilde t}

\def\tg{\tilde g}
\def\tnu{\tilde\nu}
\def\tell{\tilde\ell}
\def\tq{\tilde q}
\def\tw{\widetilde W}
\def\tz{\widetilde Z}

\def\be{\begin{equation}}  
\def\ee{\end{equation}}

\begin{document}
\begin{titlepage}
\begin{flushright}
Freiburg-THEP-06/08 \\
UH-511-1091-06
\end{flushright}

\vspace{0.5cm}
\begin{center}
{\Large \bf 
Measuring Modular Weights in Mirage Unification Models at the LHC and ILC
}\\ 
\vspace{1cm} \renewcommand{\thefootnote}{\fnsymbol{footnote}}
{\large Howard Baer $^{1,2}$\footnote[1]{Email: baer@hep.fsu.edu },
Eun-Kyung Park $^2$\footnote[2]{Email: epark@hep.fsu.edu },
Xerxes Tata $^3$\footnote[3]{Email: tata@phys.hawaii.edu },
Ting T. Wang $^2$\footnote[4]{Email: tingwang@hep.fsu.edu }} \\
\vspace{1.cm} \renewcommand{\thefootnote}{\arabic{footnote}}
{\it 
1. Physikalisches Institut,
Albert Ludwigs Universit\"at, 79104 Freiburg i. Br., Germany \\
2. Dept. of Physics,
Florida State University, Tallahassee, FL 32306, USA \\
3. Dept. of Physics and Astronomy,
University of Hawaii, Honolulu, HI 96822, USA
}

\end{center}


\vspace{0.5cm}
\begin{abstract}
String compactification with 
fluxes yields MSSM soft SUSY breaking terms that receive comparable
contributions from modulus
and anomaly mediation whose relative strength is governed by 
a phenomenological parameter $\alpha$.
Gaugino and first/second generation 
(and sometimes also Higgs and third generation) scalar mass parameters
unify at a mirage unification scale $Q \not= M_{\rm GUT}$, determined
by the value of $\alpha$.
The ratio of scalar to gaugino masses at this mirage
unification scale depends directly on the scalar field modular weights, 
which are fixed in turn by the brane or brane intersections on which the 
MSSM fields
are localized. We outline a program of measurements which can in principle 
be made at the CERN LHC 
and the International Linear $e^+e^-$ collider (ILC) which can lead to a 
determination of the modular weights.

\vspace{0.5cm}

\end{abstract}


\end{titlepage}

Superstring theory provides a consistent
quantum theory of gravity, 
together with
all the necessary ingredients for a theory that potentially unifies all
four forces of nature. However, in order to make any contact with
phenomenology, it is essential to understand how the degeneracy
associated with the many flat directions in the space of
scalar fields (the moduli) is lifted to yield the true ground state, 
since many quantities relevant for physics at accessible energies are
determined by the ground state values of these moduli. 
The discovery of a new class of compactifications, where the extra spatial
dimensions are curled up to small sizes with fluxes of additional fields
trapped along these extra dimensions 
has been exploited by
Kachru {\it et al.} (KKLT)\cite{kklt} to construct
a concrete model with a stable, calculable
ground state with a positive cosmological constant and broken supersymmetry.
This toy model is
based on type-IIB superstrings including compactification 
with fluxes to a Calabi-Yau orientifold.
While the background fluxes serve to stabilize the dilaton and the moduli
that determine the shape of the compact manifold, it is necessary to 
invoke
a non-perturbative mechanism such as gaugino condensation on a $D7$
brane to stabilize the size of the compact manifold.  
Finally, a
non-supersymmetric anti-brane ($\overline{D3}$) 
is included in order to break supersymmmetry and obtain a de Sitter
universe as required by observations. 
The resulting low energy theory thus has no unwanted light moduli, has a
broken supersymmetry, and a positive cosmological constant, but of
course does not yield the Standard Model (SM). The existence of these flux
compactifications with stable calculable minima having many desired 
properties may be viewed as
a starting point for the program of discovering a string ground
state that may lead to the (supersymmetric) Standard Model at low
energies, and which is consistent with various constraints from
cosmology.

These considerations have 
recently motivated several authors to analyze 
the structure of the soft SUSY breaking (SSB) terms in models based on 
a generalization of the KKLT
set-up \cite{choi}.  The key observation is that because
of the mass hierarchy,
\begin{equation}
m_{\rm moduli}\gg m_{3/2}\gg m_{\rm SUSY} ,
\label{eq:hierarchy}
\end{equation}
that develops in these models,
these terms receive comparable contributions via both
modulus (gravity) and anomaly mediation of SUSY breaking\cite{amsb}, 
with their relative size 
parametrized by one new 
parameter $\alpha$. 
Moreover, the  hierarchy (\ref{eq:hierarchy}) that leads to
this mixed modulus-anomaly mediated SUSY breaking (MM-AMSB) automatically
alleviates  phenomenological problems from late
decaying moduli and gravitinos that could disrupt, for instance, 
the predictions of light element abundances from Big Bang nucleosynthesis.
Upon integrating out the heavy dilaton field
and the shape moduli, we are left with an effective broken supergravity
theory of the observable sector fields denoted by $\hat{Q}$ and
the size modulus field $\hat{T}$. 
The K\"ahler
potential depends on the location of matter and Higgs superfields in the
extra dimensions via their modular weights $n_i = 0 \ (1)$ for matter
fields located on $D7$ ($D3$) branes, or $n_i=1/2$ for chiral multiplets
on brane intersections, while the gauge kinetic function
$f_a={\hat{T}}^{l_a}$, where $a$ labels the gauge group, is
determined by the corresponding location of the gauge supermultiplets,
since the power $l_a= 1 \ (0)$ for gauge fields on $D7$ ($D3$)
branes \cite{choi3}.


Within the MM-AMSB model,
the SSB gaugino mass parameters, trilinear SSB
parameters and sfermion mass parameters, all renormalized just below the
unification scale (taken to be $Q=M_{\rm GUT}$), are given by,
\begin{eqnarray}
M_a&=& M_s\left( l_a \alpha +b_a g_a^2\right),\label{eq:M}\\
A_{ijk}&=& M_s \left( -a_{ijk}\alpha +\gamma_i +\gamma_j +\gamma_k\right),
\label{eq:A}\\
m_i^2 &=& M_s^2\left( c_i\alpha^2 +4\alpha \xi_i -
\dot{\gamma}_i\right) ,\label{eq:m2}
\end{eqnarray}
where $M_s\equiv\frac{m_{3/2}}{16\pi^2}$,
$b_a$ are the gauge $\beta$ function coefficients for gauge group $a$ and 
$g_a$ are the corresponding gauge couplings. The coefficients that
appear in (\ref{eq:M})--(\ref{eq:m2}) are given by
$c_i =1-n_i$, $a_{ijk}=3-n_i-n_j-n_k$ and
$\xi_i=\sum_{j,k}a_{ijk}{y_{ijk}^2 \over 4} - \sum_a l_a g_a^2
C_2^a(f_i).$ 
Finally, $y_{ijk}$ are the superpotential Yukawa couplings,
$C_2^a$ is the quadratic Casimir for the a$^{th}$ gauge group
corresponding to the representation to which the sfermion $\tf_i$ belongs,
$\gamma_i$ is the anomalous dimension and
$\dot{\gamma}_i =8\pi^2\frac{\partial\gamma_i}{\partial \log\mu}$.
Expressions for the last two quantities involving the 
anomalous dimensions can be found in the Appendix of Ref. \cite{flm}.

The MM-AMSB model is completely specified by the parameter set,
\begin{equation}
\ m_{3/2},\alpha ,\ \tan\beta ,\ sign(\mu ),\ n_i,\ l_a.
\end{equation}
The mass scale for the SSB parameters is dictated by the gravitino mass
$m_{3/2}$. The phenomenological parameter $\alpha$,
which could be of either sign, determines the relative contributions of
anomaly mediation and gravity mediation to the soft terms, and as
mentioned above $|\alpha| \sim {\cal O}(1)$ is the hallmark
of this scenario. Non-observation of large flavor changing neutral
currents implies common modular weights of particles with the same 
gauge quantum numbers. Grand Unification implies matter particles
within the same GUT multiplet have common modular weights, and that
the $l_a$ are universal. We will assume that all $l_a=l$ and, 
for simplicity, a common modular weight for all matter particles,
but allow a different (common) one for the two Higgs doublets of the MSSM.
The main purpose of this analysis is to see to what extent it will be
possible to confirm our assumptions and deduce the value of $l$ and the
modular weights, assuming that SUSY is discovered at the LHC and is
further studied at a TeV $e^+e^-$ linear collider.  
Other aspects of MM-AMSB phenomenology have been examined in the
literature \cite{choi3,flm,eyy,kn,bptw}.

The universality of the $l_a$
leads to the
phenomenon of {\it mirage unification}\cite{choi3,flm} of gaugino
masses. In other words, gaugino mass
parameters $M_i$ 
(assuming that these can be extracted from the data) when extrapolated 
using one loop renormalization group equations (RGEs) would unify
at a scale $Q=\mu_{\rm mir} \not= M_{\rm GUT}$, the scale of unification
of gauge couplings. Indeed, the observation of gaugino unification at
the mirage unification scale,
\be
\mu_{\rm mir}=M_{\rm GUT}e^{-8\pi^2/(l \alpha)} ,
\ee
would strikingly point to such a scenario. If $\alpha < 0$, $\mu_{\rm
  mir}> M_{\rm GUT}$, though one would have to continue extrapolation
  using MSSM RGEs to discover this! We assume here that $l
\not=0$, since this would be distinguished by a gaugino mass pattern as in
the AMSB framework. 
While $\mu_{\rm mir}$ determines $l\alpha$, the (unified) value of the 
the gaugino masses extrapolated to $Q=\mu_{\rm mir}$ is $M_a(\mu_{\rm
  mir})= M_s \times (l\alpha)$, and so gives the value of $M_s$ (and so
$m_{3/2}$). 

\begin{figure*}[htbp]
\begin{center}
\epsfig{file=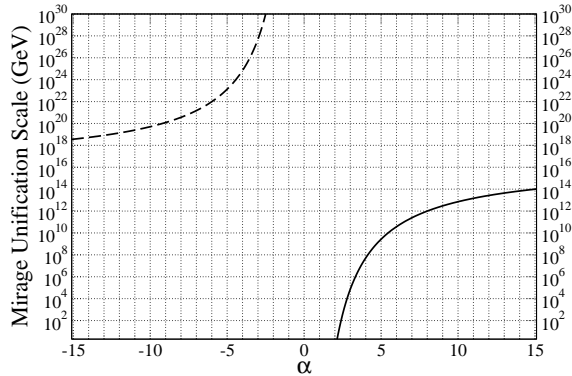,height=8cm,angle=-90}
\end{center}
\caption{A plot of the mirage unification scale versus
modulus-AMSB mixing parameter $\alpha$, assuming $l=1$.
\label{fig:1}}
\end{figure*}

We show the mirage unification scale versus $l\alpha$ in
Fig. \ref{fig:1} for $l=1$. 
The existence of a mirage unification scale is taken to be a 
``smoking gun'' signature for MM-AMSB models. 
If supersymmetry is discovered and the various soft parameters 
are precisely measured at the weak scale, then extrapolation of the soft
parameters via the RGEs to a point of unification\cite{zerwas}
at a scale $\mu_{\rm mir} \not= M_{\rm GUT}$
would indicate that nature is in fact described by a MM-AMSB model
with mirage unification!
In the process, the scale $\mu_{\rm mir}$, or equivalently $l\alpha$, would be 
measured.

In the MM-AMSB framework with universal matter modular weights (for the first
two generations whose Yukawa couplings are negligible), the SSB matter
mass parameters also unify at $Q=\mu_{\rm mir}$, with $m_i^2(\mu_{\rm
  mir})= M_s^2 c_i \alpha^2$. If the extrapolated values of selectron or 
first generation squark mass parameters indeed converge at the same
unification scale as gaugino parameters, it would provide striking 
confirmation of this framework. 
Taking the ratio of first/second generation
scalar to gaugino mass parameters yields,
\be
\left.\frac{m_i}{M_a}\right|_{\mu_{\rm mir}} ={\sqrt{c_i}\over l} .
\label{ratio:1}
\ee

The obvious question is whether it is possible to disentangle the values
of $c_i$ and $l$. A look at the boundary conditions for the gaugino and
first/second generation SSB parameters shows that these depend only on
the combinations $M_s$, $l\alpha$ and $c_i/l^2$: this is obvious for
the gaugino masses, while for the scalar masses, this is clearly also
the case since $\xi_i \propto l$ as long as the Yukawa couplings are
negligible. Thus {\it it is impossible even in principle to disentangle
$c_i$ and $l$ from these measurements alone}. To do so, even in
principle, it is essential to determine either the SSB third generation
mass parameters or the $A$-parameters. While it is clear that the
boundary condition (\ref{eq:A}) depends on $c_i/l$ (together with
$l\alpha$ and $M_s$), it is not difficult to check that the Yukawa
coupling terms in $\xi_i$ also depend on $c_i/l$. A precise
determination of third generation SSB or of $A$ parameters would, in
principle, allow us to separately obtain $l$ and thus check whether or
not this is unity. This may well be possible via a study of the stau
sector at an electron-positron collider\cite{mihoko}, and perhaps, via the
stop sector if $e^+e^- \to \tst_1\bar{\tst}_1$ pairs is accessible. We will
not examine this any further but assume that $l=1$ for the remainder of
this paper.

%
%

In this case it is clear that the matter modular weights can be determined from
(\ref{ratio:1}) once SSB scalar and gaugino mass parameters are 
determined.  What would
it take to measure these? It would likely take a combination of
measurements from the CERN LHC and a linear $e^+e^-$ collider such as
the proposed International Linear Collider (ILC), which would operate at
CM energies of around $\sqrt{s}\sim 0.5-1$ TeV, and/or by the CERN CLIC
linear collider, which is proposed to operate in the multi-TeV regime
\cite{georg}.

The weak scale gaugino mass $M_3$ at tree level is the same as the
gluino mass $m_{\tg}$, although the relation between these quantities
gets corrected by known loop effects that give corrections up to
$\sim$30\%\cite{mgl} (though in the present framework, we do not expect
very large corrections because the ratio $m_{\tq}/m_{\tg}$ is not
especially large).  The gluino mass has been shown to be measureable at
the LHC in several benchmark cases via $\tg\tg$ production followed by
gluino cascade decays\cite{frank}.  Another measurement LHC can make is
the $m_{\tz_2}-m_{\tz_1}$ mass difference if
$\tz_2\to\tz_1\ell\bar{\ell}$ decays occur at a sufficient rate\cite{mll}. 
In this case, the dilepton invariant mass distribution will offer one
strong constraint on the neutralino mass matrix, which depends on the
gaugino masses $M_1$ and $M_2$, as well as on the superpotential Higgs
mass term $\mu$ and the ratio of Higgs vevs $\tan\beta$. Moreover, from
the shape of the end-point of the $m_{\ell\ell}$ spectrum it may be
possible to determine whether or not the higgsino component of the
neutralinos is large or small, at least in the case that $M_1/M_2 > 0$
at the weak scale \cite{kn}: for very small higgsino components,
$m_{\tz_2}-m_{\tz_1}=M_2-M_1$.

The gaugino masses $M_1$ and $M_2$, and possibly
the parameter $\mu$, may be extracted
at a LC by a combination of measurements of $\tw_1^+\tw_1^-$
production, $\tz_1\tz_2$ production and $\tw_1^\pm\tw_2^\mp$ 
production\cite{jlc,bmt,tadas,tesla}. 
While really a measurement of only two of the three 
SSB gaugino masses is necessary
to establish the value of $\mu_{\rm mir}$ and $M_a(\mu_{\rm mir})$, the measurement
and extrapolation of the third gaugino mass would offer striking support
for a
mirage unification hypothesis.

Turning to matter scalar masses, the CERN LHC has some ability to
measure squark masses, at least in some benchmark studies\cite{frank},
although it will be difficult to tell the flavour or type of squark
being produced.  It may also be possible for LHC to extract some 
information on slepton masses, not so much from direct slepton
production\cite{slep} as much as from their production in cascade
decays in fortituous cases, 
or via their influence on the shape of the
dilepton invariant mass spectrum from $\tz_2\to\tz_1\ell\bar{\ell}$ or
$\tz_2\to\ell\tell$ decays\cite{shape}.

For a LC, the first and second generation $\tell_R$, $\tell_L$ and
$\tnu_{\ell}$ masses should be readily measured if pair production of
these scalars is allowed either through the lepton energy spectrum
endpoints\cite{jlc,bmt} or via threshold measurements \cite{tesla}. In
addition, if squark pair production is accessible, then squark masses
should be measureable to some degree, along with squark type, using the
beam polarization tool \cite{ff}. Again, only
two scalar masses (such as $m_{\tell_L}$ and $m_{\tell_R}$) need be
measureable to establish mirage unification at $\mu_{\rm mir}$ (which should
coincide with the unification scale obtained via gauginos) and the
associated soft term masses at $\mu_{\rm mir}$ that can yield information
about the corresponding modular weights, and also serve to test our
hypothesis that the modular weights are the same for all matter particles.

It would be interesting to be able to check that $l=1$. As discussed
above, this entails a determination of either the $A$-parameters or
third generation SSB masses whose evolution receives sizeable
contributions from Yukawa couplings. This appears to be very difficult
at the LHC, though in some fortituous cases where $\tb_1$ is light
enough to be produced in gluino cascade decays some information may be
possible \cite{frank}. Stau production at the ILC may offer the best
access to the third generation parameters since, at least in favourable
cases, the mass as well as the stau mixing angle may be determined
\cite{mihoko}. Unfortunately, unless $\tan\beta$ is also large, the
effects of the Yukawa couplings that are essential for separating out
the value of $l$ will be small. Information about $l$ can presumably be
obtained via a study of $t$-squark system at an electron-positron
collider with sufficiently high energy, {\it but only if $n_H$ can be
obtained via measurements in the Higgs sector.}

Higgs scalar SSB mass parameters appear to be especially interesting
because these can potentially be used to both determine the modular
weights in the Higgs sector, and to obtain information on $l$ (since
their boundary condition depends also on the Yukawa couplings). These may be
extracted at a linear collider if the heavy neutral and/or charged Higgs
bosons are accessible. We note that one of the tree level MSSM scalar
potential minimization conditions reads $\mu^2
=\frac{m_{H_d}^2-m_{H_u}^2\tan^2\beta}{(\tan^2\beta -1)}
-\frac{M_Z^2}{2}$ while the pseudoscalar Higgs mass $m_A$ is given by
$m_A^2=m_{H_u}^2+m_{H_d}^2+2\mu^2$, so that in principle a determination
of $\mu$, $m_A$ and $\tan\beta$ would determine these quantities. Of
course, these tree level relations suffer important loop corrections
that depend on other sparticle masses, which would have to be taken into
account.  A variety of cases have been investigated at both the LHC and
the ILC for measuring the heavier Higgs boson masses and the parameter
$\tan\beta$ \cite{gunion,georg}, and as noted earlier, $\mu$ should be
extractable at an ILC especially if $\tw_1^\pm\tw_2^\mp$ production is
accessible. The extraction of Higgs modular weights is, however, more
complicated than for first/second generation matter since, because of
Yukawa coupling effects, the weak scale values of $m_{H_u}^2$ and
$m_{H_d}^2$ are not expected to extrapolate (via one loop evolution) to
a common value at $Q=\mu_{\rm mir}$ except for the special cases~3 and 8
in Table~\ref{tab:1} below; for these special cases, (\ref{eq:m2})
applies, and the value of $m_{H_i}^2(\mu_{\rm mir})$ yields
$\sqrt{c_H}/l$. In principle, the GUT scale value of the Higgs SSB
parameters depend on $c_i/l$ so it is possible that if these can be
determined to a sufficiently good precision, these can be used to
extract the value of $l$, and check that this is consistent with that
obtained via a study of staus or top squarks. For the other cases in
Table~\ref{tab:1}, the extraction of $n_H$ seems more
difficult.\footnote{If we assume $l=1$, and assume a universal value of
$n_{\rm matter}$, it should be possible to extract $n_{H}$ by
extrapolating the Higgs SSB mass parameters to the GUT scale with
sufficient precision. Since this requires a knowledge of many masses and
their mixings, we do not make any representation that this can be done
in practice.}



\begin{figure*}[htbp]
\begin{center}
\epsfig{file=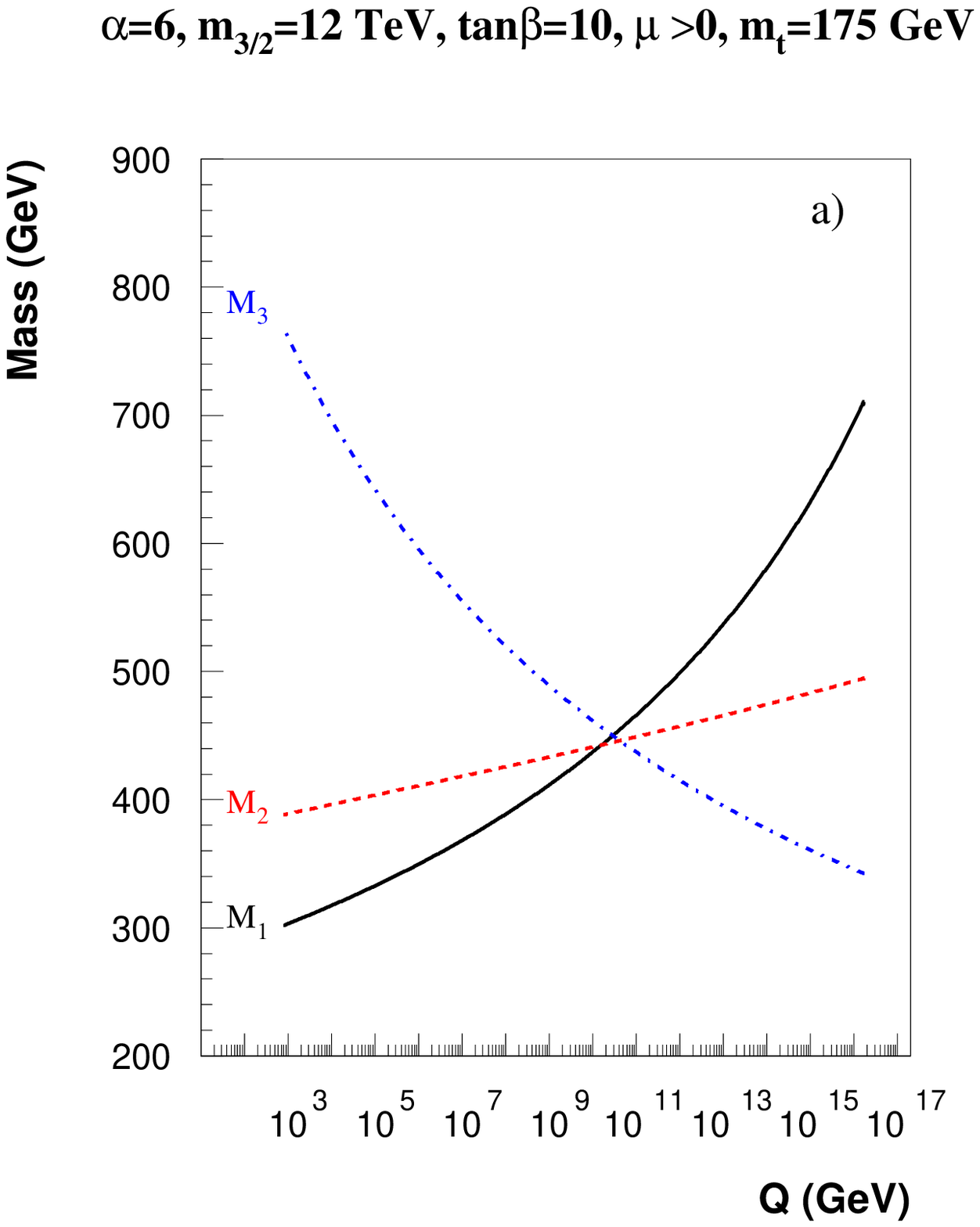,width=6cm} 
\epsfig{file=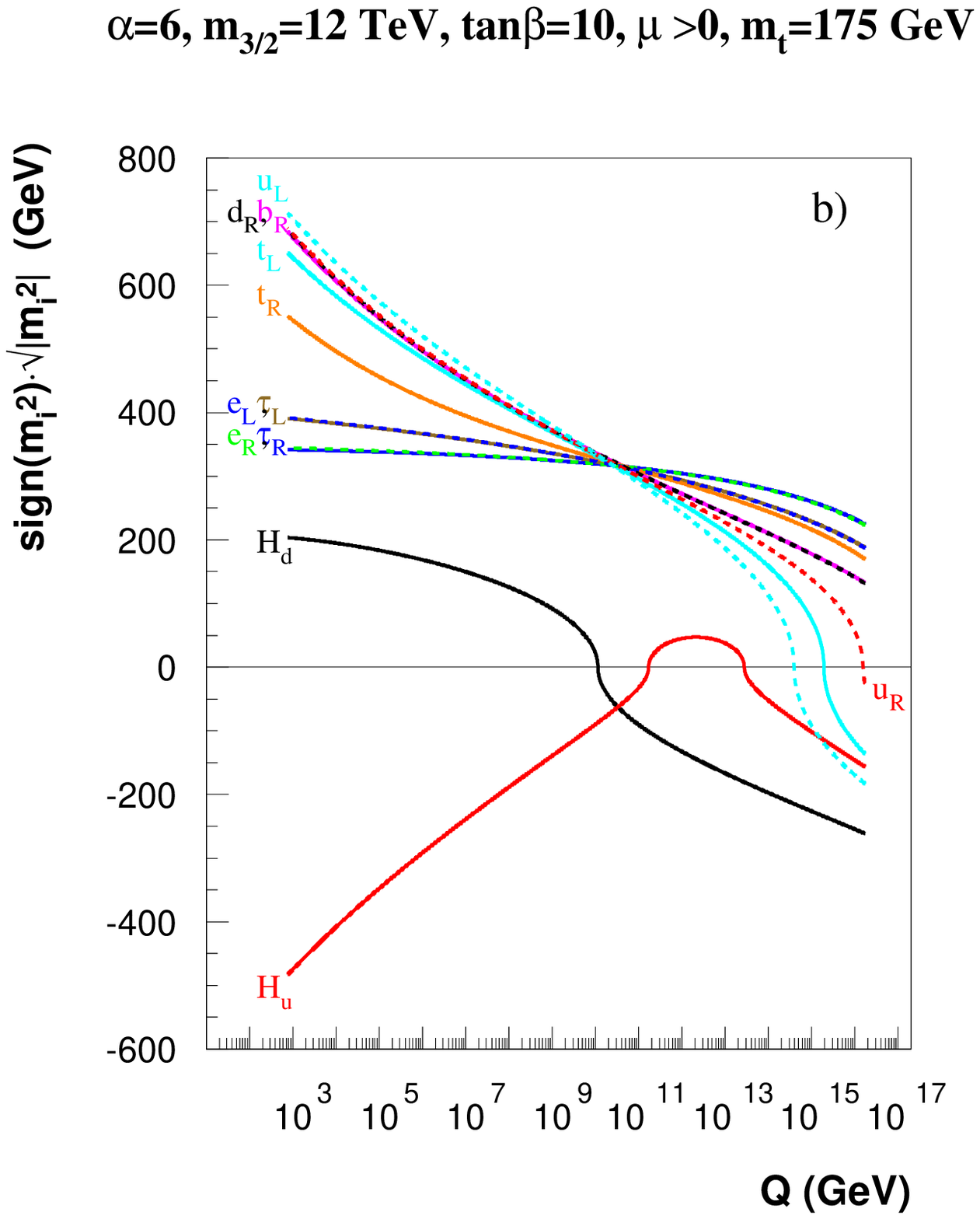,width=6cm} 
\end{center}
\caption{
Evolution of {\it a}) the gaugino masses $M_1$, $M_2$ and $M_3$ from
$Q=M_{\rm weak}$ to $Q=M_{\rm GUT}$ in the MM-AMSB model for $\alpha =6$,
$m_{3/2}= 12$ TeV, $\tan\beta =10$, $\mu >0$ and $m_t=175$ GeV and
for $l_a=1$, $n_{\rm matter}={1\over 2}$ and $n_H=1$.
In frame {\it b}, we show evolution of
scalar soft masses from $Q=M_{\rm weak}$ to $Q=M_{\rm GUT}$ for
the same parameter choices. Whereas the unification of gaugino and first
generation sfermion mass parameters is quite generic, the unification of  
the corresponding Higgs and third generation mass parameters is special
to our choice of modular weights as discussed in the text.}
\label{fig:2}
\end{figure*}

We illustrate in Fig. \ref{fig:2}{\it a}) the gaugino mass 
unification in an MM-AMSB model with
$\alpha =6$, $m_{3/2}=12$ TeV, $\tan\beta =10$ and $\mu >0$ for 
$m_t=175$ GeV, $n_{\rm matter}={1\over 2}$ and $n_H=1$. 
It is apparent that $\mu_{\rm mir}\sim 10^{11}$ GeV,
while $M_a(\mu_{\rm mir})\sim 450$ GeV. In Fig. \ref{fig:2}{\it b}), we show
the evolution of various matter and Higgs scalar soft masses from
$M_{\rm weak}$
to $M_{\rm GUT}$. The soft parameters again unify at $\sim 10^{11}$ GeV, 
while matter scalars have a mass $\sim 320$ GeV and Higgs scalars have a mass
$\sim 0$ GeV. We have checked that in fact the Higgs masses evolve to zero
at $Q=\mu_{\rm mir}$ if one-loop RGEs are used, so that the off-set of
$m_{H_{u,d}}$ at $Q=\mu_{\rm mir}$ is a consequence of the two-loop RGEs
that are inherent in Isajet, which we use for our calculation of sparticle
masses\cite{isajet}. We stress that first/second generation masses
always unify at $\mu_{\rm mir}$, while the unification of third
generation and Higgs SSB mass parameters is special to the choice of
modular weights.

In Table \ref{tab:1}, we show nine cases of matter and Higgs field
modular weights. It is clear that a determination of matter modular
weights from masses of gauginos and first generation sfermions at future
colliders will localize the models in one of three groups where
$m_i/M_a$ at the unification scale is 0 (cases~3, 6, 9) , $1/\sqrt{2}$
(cases~2, 5, 8) or 1 (cases~1, 4, 7). Information about third
generation, Higgs or trilinear SSB parameters will be essential to
further separate the degeneracies. If, for instance, third generation
mass parameters also unify at $\mu_{\rm mir}$, we will know that we are
in cases 3 or 8. In other cases, more careful scrutiny will be necessary
since, for example, the distinction between cases 6 and 9 is only
possible via the value of the Higgs or trilinear SSB parameters. As
mentioned above, this may be possible if we assume $l=1$ to deterime
$n_H$. If we can extrapolate the weak scale $A$-parameters to $M_{\rm
GUT}$, we can then test the consistency of this assumption: like the
extrapolation of SSB Higgs mass parameters, this requires us to know
masses and mixings of many sparticles, and detailed studies are needed
to decide whether the extrapolation \cite{zerwas} to $M_{\rm GUT}$ can
be done with the required precision.  In the special cases~3 and 8, a
complete determination of the modular weights along with the value of
$l$ appears to be possible by combining the data from the LHC with that
from an electron positron collider.

%
\begin{table}[hbt]
\begin{center}
\begin{tabular}{lccccccccc}
\hline
case & 1 & 2 & 3 & 4 & 5 & 6 & 7 & 8 & 9 \\
\hline
$n_H$ & 0 & 0 & 0 & $1/2$ & $1/2$ & $1/2$ & 1 & 1 & 1 \\
$n_{matter}$ & 0 & $1/2$ & 1 & 0 & $1/2$ & 1 & 0 & $1/2$ & 1 \\
\hline
\end{tabular}
\end{center}
\caption{ Nine cases of Higgs and matter modular weights which are
explored in the text.
}
\label{tab:1}
\end{table}
%

%
%

To summarize, 
in supersymmetric models with a KKLT type vacuum, SSB terms receive
comparable contributions from modulus and anomaly mediated SUSY breaking
resulting in the phenomenon of mirage unification. The mirage
unification scale should be measureable by extrapolation of soft SUSY
breaking masses from $Q=M_{\rm weak}$ to $Q=\mu_{\rm mir}$ via one loop
RGEs.  The ratio of first/second generation soft masses 
to gaugino masses at the mirage
unification scale offers a direct measurement of the scalar field
modular weights which, in turn, provides information about
the dimensionality of the branes on which these scalar fields reside.


This research was supported in part by grants 
from the United States
Department of Energy.


\end{document}